\documentclass[pre,twocolumn,groupedaddress,showpacs,showkeys,amsmath]{revtex4}

\usepackage{graphicx}
\usepackage{dcolumn}
\usepackage{bm}

\begin{document}


\title{The universality class of the continuous phase transition in the $2D$ ``Touch and Stop'' cluster growth percolation model}

\author{O. Melchert}
\email{oliver.melchert@uni-oldenburg.de}
\affiliation{
Institut f\"ur Physik, Carl von Ossietzky Universit\"at Oldenburg, D-26111 Oldenburg, Germany\\
}

\date{\today}


\begin{abstract}
We consider the ``Touch and Stop'' cluster growth percolation (CGP) model on the two dimensional square lattice.
A key-parameter in the model is the fraction $p$ of occupied ``seed'' sites that act as nucleation centers
from which a particular cluster growth procedure is started. Here, we consider two growth-styles: rhombic and 
disk-shaped cluster growth. For intermediate values of $p$ the final 
state, attained by the growth procedure, exhibits a cluster of occupied sites that spans the entire lattice. 
Using numerical simulations we investigate the percolation probability and the order parameter 
and perform a finite-size scaling analysis for lattices of side length up to $L\!=\!1024$ in 
order to carefully determine the critical exponents that govern the respective transition.
In contrast to previous studies, reported in [Tsakiris \emph{et al.}, Phys.\ Rev.\ E 82 (2010) 041108], 
we find strong numerical evidence that the CGP model is in the standard percolation universality class.
\end{abstract} 

\maketitle

\section{Introduction \label{sect:introduction}}

The pivotal question in standard percolation \cite{stauffer1979,stauffer1994} is that 
of connectivity. A basic example is $2D$ random site percolation, where 
one studies a lattice in which a random fraction $p$ of the 
sites is ``occupied''.
Clusters composed of adjacent occupied sites are then analyzed 
regarding their geometric properties. 
Depending on the fraction $p$ of occupied sites, the geometric 
properties of the clusters change, leading from a phase 
with rather small and disconnected clusters to a phase, where there is 
basically one large cluster covering the lattice.  
Therein, the appearance of an infinite, i.e.\ percolating, cluster is 
described by a second-order phase transition.  

There is a wealth of literature on a multitude of variants on the
above basic percolation problem that model all kinds of phenomena,
ranging from simple configurational statistics to ``string''-bearing 
models that also involve a high degree of optimization, e.g.\ describing
vortices in high $T_c$ superconductivity \cite{pfeiffer2002,pfeiffer2003}
and domain wall excitations in disordered media such as $2D$ spin glasses \cite{cieplak1994,melchert2007} and 
the $2D$ solid-on-solid model \cite{schwarz2009}. 
Besides discrete lattice models there is also interest in studying 
continuum percolation models, where recent studies reported on highly
precise estimates of critical properties for spatially
extended, randomly oriented and possibly overlapping objects with various 
shapes \cite{MertensMoore2012}.

\begin{figure}[t!]
\centerline{
\includegraphics[width=1.0\linewidth]{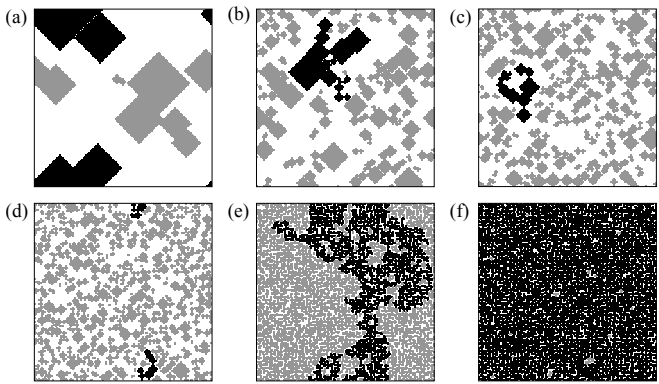}}
\caption{
Samples of final configurations obtained for the CGP model 
on $2D$ square lattices of side length $L\!=\!128$.
The snapshots relate to different values of the
initial density of active seed sites:
$p\!=\!0.001$, $0.01$, $0.02$, $0.05$, $0.495$, $0.7$, for 
subfigures (a--f), respectively. 
In the limit of large system sizes and above the critical point 
$p_c\approx 0.495$, the largest cluster in the final configuration spans the 
lattice along at least one direction. In each subfigure, the largest
cluster is colored black and all finite clusters are colored grey.
\label{fig:2Dsamples}}
\end{figure}

One such variant of the above basic percolation model is the
recently proposed ``Touch and Stop'' cluster growth percolation (CGP)
model \cite{Tsakiris2010a,Tsakiris2010b}.
In the CGP model, a random fraction of $p$ sites is distinguished 
to comprise a set of ``seed'' sites for which a particular cluster 
growth procedure is evolved (see sect.\ \ref{sect:model}). 
I.e., starting from the seed sites, clusters are grown by assimilating all
nearest neighbor sites layer by layer in an iterative, discrete time fashion. 
As soon as one cluster comes in contact with other clusters, the growth procedure 
for all involved clusters is stopped.
When there is no growing cluster left, the cluster growth procedure is completed and 
the connected regions of adjacent occupied sites, i.e.\ the final clusters, are analyzed regarding
their geometric properties.
To support intuition: if $p$ is rather small, the growth of a particular cluster
is unlikely to be hindered by other clusters within the first few time steps, 
since the typical distance between the respective seed sites is large compared to the lattice spacing.
Consequently, the final clusters are rather large, see Fig.\ \ref{fig:2Dsamples}(a).
As $p$ increases, the typical size of clusters in the final configuration decreases due to the 
increasing density of initial seed sites, see Figs.\ \ref{fig:2Dsamples}(a--c). 
Increasing $p$ even further leads to an increasing size of the largest cluster.
This is due to the larger probability to yield adjacent seed sites already in the initial configuration, 
for which a growth procedure is subsequently inhibited. For increasing $p$
and due to spacial homogeneity this in turn leads to an increase of the size of the largest cluster,
see Figs.\ \ref{fig:2Dsamples}(c--e).

The two distinct scaling regimes of the largest cluster size were discussed in Ref.\ \cite{Tsakiris2010a}.
The behavior at small values of $p$ was illustrated to be a finite size effect and the scaling at 
intermediate values of $p$ was found to signal a continuous phase transition qualitatively similar to 
the standard percolation transition.
In Ref.\ \cite{Tsakiris2010b}, the critical properties of the latter transition were addressed. 
Albeit the authors of Ref.\ \cite{Tsakiris2010b} perform a scaling analysis 
to estimate the critical point and a set of critical exponents that 
govern the phase transition, we are convinced that the analysis can be improved in various ways.
In this regard, we here revisit the continuous phase transition found for the CGP model
at intermediate values of $p$ and perform a finite size scaling analysis considering 
geometric properties of the largest cluster for the individual final configurations
obtained using computer simulations.
To this end, the results we report in sect.\ \ref{sect:results} below (obtained using two 
different cluster growth styles, i.e.\ rhombic and disk-shaped cluster growth) are different from those
presented in Ref.\ \cite{Tsakiris2010b} in the following respect: in contrast to their findings we here 
present strong numerical evidence that the percolation transition of the CGP model at intermediate values
of $p$ is in the same universality class as standard percolation.
By employing the data collapse technique for different observables we aim
to yield maximally justifiable results due to a high degree of numerical redundancy. 
Further, we analyze the distribution of cluster sizes right at the critical point in order to 
numerically assess a particular critical exponent that was previously only computed via 
scaling relations.

The remainder of the presented article is organized as follows.
In section \ref{sect:model}, we introduce the model in 
more detail.
In section \ref{sect:results}, we outline the 
data collapse technique and we list the results of 
our numerical simulations. Therein, the discussion is
focused on rhombic cluster growth. Section \ref{sect:conclusions} 
concludes with a summary.
Appendix \ref{ap:appendixA} shows a further analysis of the
order parameter resembling the one presented in Ref.\ \cite{Tsakiris2010b},
and appendix \ref{ap:appendixB} elaborates on the 
results obtained for disk-shaped cluster growth.

\section{Model and Algorithm\label{sect:model}}

Here, we consider $2D$ square lattices of side length $L$ and $N=L^2$ sites.
Initially, a starting configuration is prepared by occupying a random fraction 
$p$ of the lattice sites via \emph{seed} sites. These seeds indicate the centers 
of a set of \emph{active clusters} that will evolve during the growth procedure. 
The remaining sites are considered \emph{empty}.
For the set of active clusters, the growth procedure consists of the repeated 
application of the following two steps: 
(i) to perform a single sweep, consider the still active clusters sequentially in random order. For each cluster perform 
the \emph{cluster update} below.
(ii) so as to perform a single cluster update, consider the \emph{surface} of the respective cluster, 
i.e.\ the set of nearest neighbor sites of those sites that build up the 
current cluster and that do not belong to the cluster, yet. To complete a cluster update, the ``Touch and Stop'' CGP model 
discriminates the following two situations:
(ii.1) if alien sites, i.e.\ sites belonging to clusters different from 
the current one are found among the surface sites, the cluster growth procedure for 
all involved clusters is stopped. The respective clusters are further deleted from the set 
of active clusters;
(ii.2) if all surface sites are empty, amend the current cluster by the set of surface sites.
The above two steps (i) and (ii) are iterated until no active cluster is left and the final configuration 
is reached. I.e., the evolution of the system proceeds in discrete time-steps, where within one
particular time-step each active cluster is updated once. 
Note that the precise configuration of occupied sites in the final configuration might
depend on the order in which active clusters are picked in step (i).
\begin{figure}[b!]
\centerline{
\includegraphics[width=1.0\linewidth]{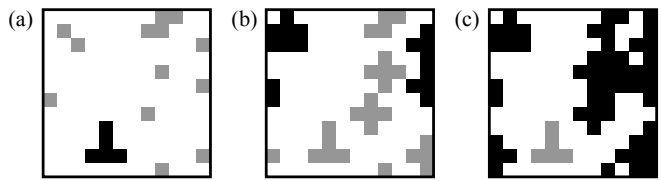}}
\caption{
Example of the cluster growth procedure for a $2D$ square lattice
of side length $L\!=\!12$. For the initial configuration of 
active seed sites shown in subfigure (a), the growth procedure
is completed after the two steps (a--b) and (b--c), already.
Subfigure (c) indicates the final configuration. See text for more details.
\label{fig:algExample}}
\end{figure}  

Fig.\ \ref{fig:algExample} illustrates the cluster growth procedure for a square lattice of 
side length $L=12$, where initially $18$ seed sites are present (see Fig.\ \ref{fig:algExample}(a)), 
i.e.\ $p=0.125$. 
The $18$ seed sites are arranged in $10$ clusters. Consequently, the 
initial configuration exhibits small clusters of adjacent occupied sites for which,
according to step (ii) above, not a single cluster update is performed. 
Both of these clusters occupy $5$ sites and one of them is picked to signify the largest cluster
on the lattice. In the figure it is colored black to distinguish it from the other clusters (which are colored grey).
While sweeping over the active clusters in step (i), a successful cluster update for one cluster might hinder
the growth of another cluster that still has to be considered in the respective sweep (see, e.g.,   
the two occupied next nearest neighbor sites in the upper left corner of Fig.\ \ref{fig:algExample}(a)).
After one sweep the configuration of occupied clusters has changed, see Fig.\ \ref{fig:algExample}(b).
Now, the largest cluster comprises $15$ sites and there are only three active clusters left.
The next sweep, resulting in the final configuration shown in Fig.\ \ref{fig:algExample}(c), 
completes the growth procedure. Therein, the largest cluster occupies $53$ sites and it 
spans the lattice along the vertical direction since its projection on the independent lattice 
axis covers $12$ sites in the vertical direction.

As pointed out in the introduction and discussed in Ref.\ \cite{Tsakiris2010a},
the model features two transitions: a sharp transition at a very small value of $p$, 
which is due to finite size effects, and a continuous transition at intermediate values of $p$.
Below, we will perform simulations on square systems of finite size and 
for different values of $p$ in order to assess the critical properties of the CGP model in the 
vicinity of the critical point in $2D$.
The observables we consider are derived from the set of clusters in the final configuration,
attained after the growth process is completed.
The observables will be introduced and discussed in the subsequent section.

\section{Results \label{sect:results}}
So as to assess the critical properties of the CGP model in the 
vicinity of the critical point in $2D$ we performed simulations for square lattices 
of side length up to $L=1024$. For each system size considered, we recorded data sets at
$24$ supporting points in a $p$-range that encloses the critical point on the $p$-axis. 
Each data set comprises the set of clusters in the final configuration for 
$\approx 10^4$ individual samples.
As stressed in the introduction, we consider two different cluster growth styles:
rhombic and disk-shaped cluster growth, defining the rhombic CGP model (CGP-R) and the 
disk-shaped CGP model (CGP-D), respectively. In the remainder of this section we 
report on the results for rhombic cluster growth. The results for disk-shaped 
cluster growth are detailed in appendix \ref{ap:appendixB}.

Most of the observables we consider below can be rescaled following a common scaling assumption
(formulated below for a general observable $y(p,L)$).
This scaling assumption states that if the observable obeys scaling, it can be expressed as
\begin{eqnarray}
y(p,L)= L^{-b}~f[(p-p_c) L^a], \label{eq:scalingAssumption}
\end{eqnarray}
wherein $a$ and $b$ represent dimensionless critical exponents (or ratios thereof, see below),
$p_c$ signifies the critical point, and $f[\cdot]$ denotes an unknown scaling function.   
Following Eq.\ (\ref{eq:scalingAssumption}), data curves of the observable $y(p,L)$ recorded at 
different values of $p$ and $L$ \emph{collapse}, i.e.\ fall on top of each other, if $y(p,L) L^{b}$ 
is plotted against the combined quantity $\epsilon \equiv (p-p_c) L^a$ and if the scaling 
parameters $p_c$, $a$ and $b$ 
that enter Eq.\ (\ref{eq:scalingAssumption}) are chosen properly.    
The values of the scaling parameters that yield the best data collapse signify the numerical 
values of the critical exponents that govern the scaling behavior of the underlying observable $y(p,L)$.
In order to obtain a data collapse for a given set of data curves we here perform a
computer assisted scaling analysis, see Refs.\ \cite{houdayer2004,autoScale2009}.
The resulting numerical estimates of the critical point and the critical exponents for the $2D$ ``Touch and Stop''
model implemented using rhombic and disk-shaped cluster growth as well as the critical properties of 
the standard $2D$ site percolation model (for comparison) are listed in Tab.\ \ref{tab:tab1}.
Below, we report on the results found for different observables:

\begin{figure}[t!]
\centerline{
\includegraphics[width=1.0\linewidth]{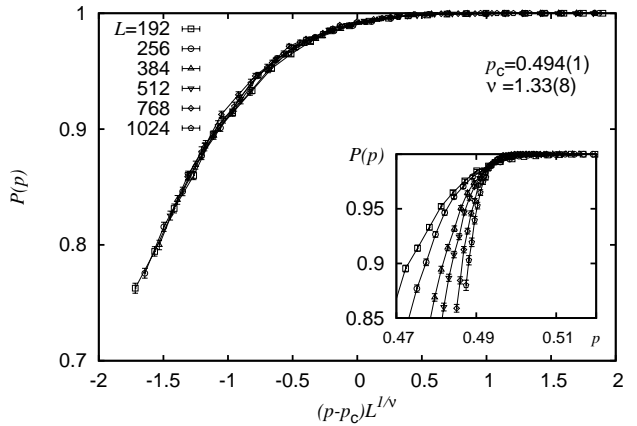}}
\caption{Finite size scaling of the spanning probability $P(p)$ 
for the CGP-R model on $2D$ square lattice of side length $L=192$ 
through $1024$, averaged over different initial seed configurations.
The main plot shows the data collapse obtained according
to Eq.\ \ref{eq:scalingAssumption}, and the inset illustrates
the raw data close to the critical point.
\label{fig:percProb_fss}} 
\end{figure}  

\begin{table}[b!]
\caption{\label{tab:tab1}
Critical properties that characterize the CGP phenomenon 
(CGP-R: rhombic clusters; CGP-D: disk-shaped clusters)
in $2D$. 
From left to right: 
Critical point $p_c$, 
critical exponents $\nu$, $\beta$, $\gamma$, and $\tau$. 
For comparison, the critical properties for the standard site percolation (SP) problem 
on the $2D$ square lattice
(figures up to the third decimal) are listed also.
} 
\begin{ruledtabular}
\begin{tabular}[c]{l@{\quad}lllllll}
 & $p_c$  & $\nu$ & $\beta$ & $\gamma$&  $\tau$ \\
\hline
CGP-R & 0.4938(7)    & 1.33(3)  & 0.141(3) & 2.38(5)  &   2.057(1) \\
CGP-D & 0.4978(5)   & 1.32(4) & 0.145(4) &  2.38(3)  &   2.050(6)  \\
SP  & 0.593         & $1.333$  & $0.139$  & $2.389$ &  $2.055$ \\
\end{tabular}
\end{ruledtabular}
\end{table}

\paragraph{Spanning probability:}
As a first observable we consider the probability $P(p)$ that the final configuration 
features a cluster that spans the lattice along at least one direction, averaged over
different initial seed configurations. It is expected to 
scale according to Eq.\ (\ref{eq:scalingAssumption}), where $a=1/\nu$ and $b=0$. 
A data collapse performed for the four largest system sizes $L=384,512,768,1024$ 
in the range $\epsilon \in [-2,2]$ on the rescaled $p$-axis yields the scaling 
parameters $p_c=0.494(1)$, and $\nu=1.34(7)$ for a quality $S=0.71$ of the data collapse, see Fig.\ \ref{fig:percProb_fss}.
As it appears, the numerical value of $\nu$ agrees well with the value for standard $2D$ percolation,
i.e.\ $\nu_{\rm perc}=4/3$.
While the location of the critical point is in agreement with the value reported in 
Ref.\ \cite{Tsakiris2010b}, the critical exponent is rather different 
(Ref.\ \cite{Tsakiris2010b} reports $\nu=1.17$). 

\begin{figure}[t!]
\begin{center}
\includegraphics[width=1.0\linewidth]{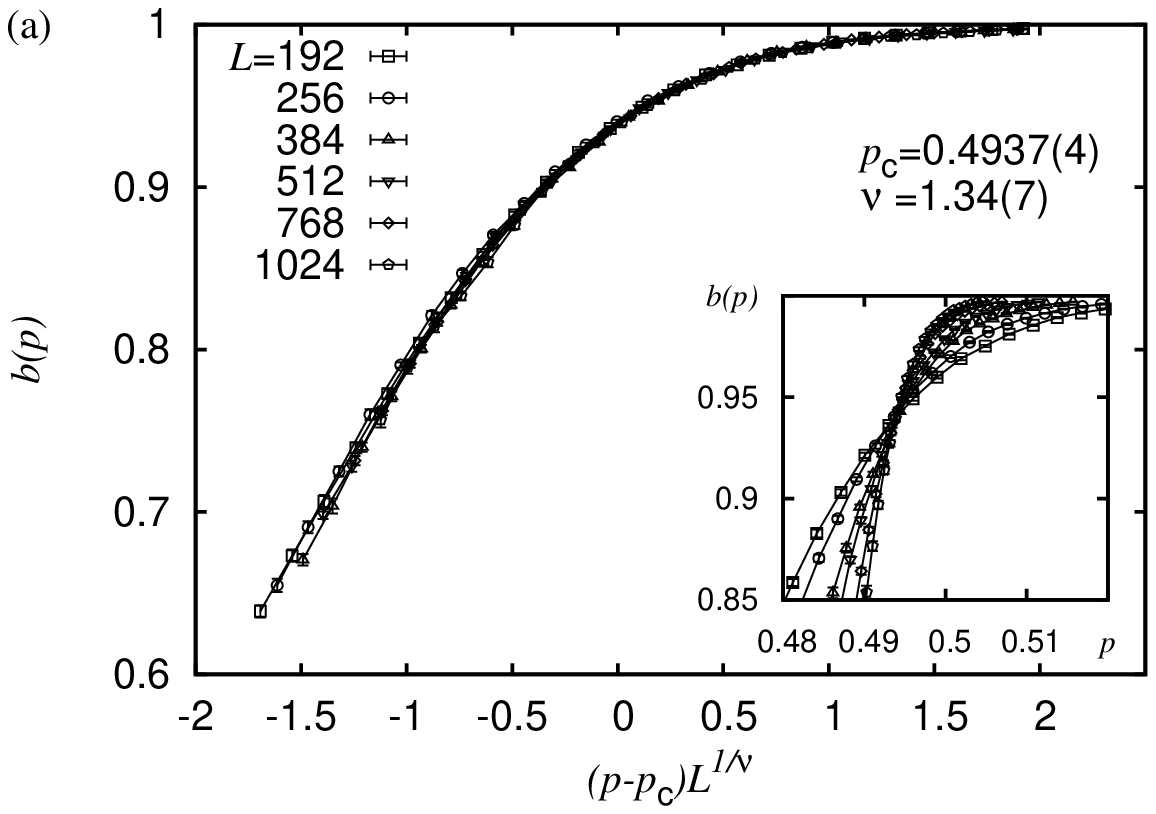}
\includegraphics[width=1.0\linewidth]{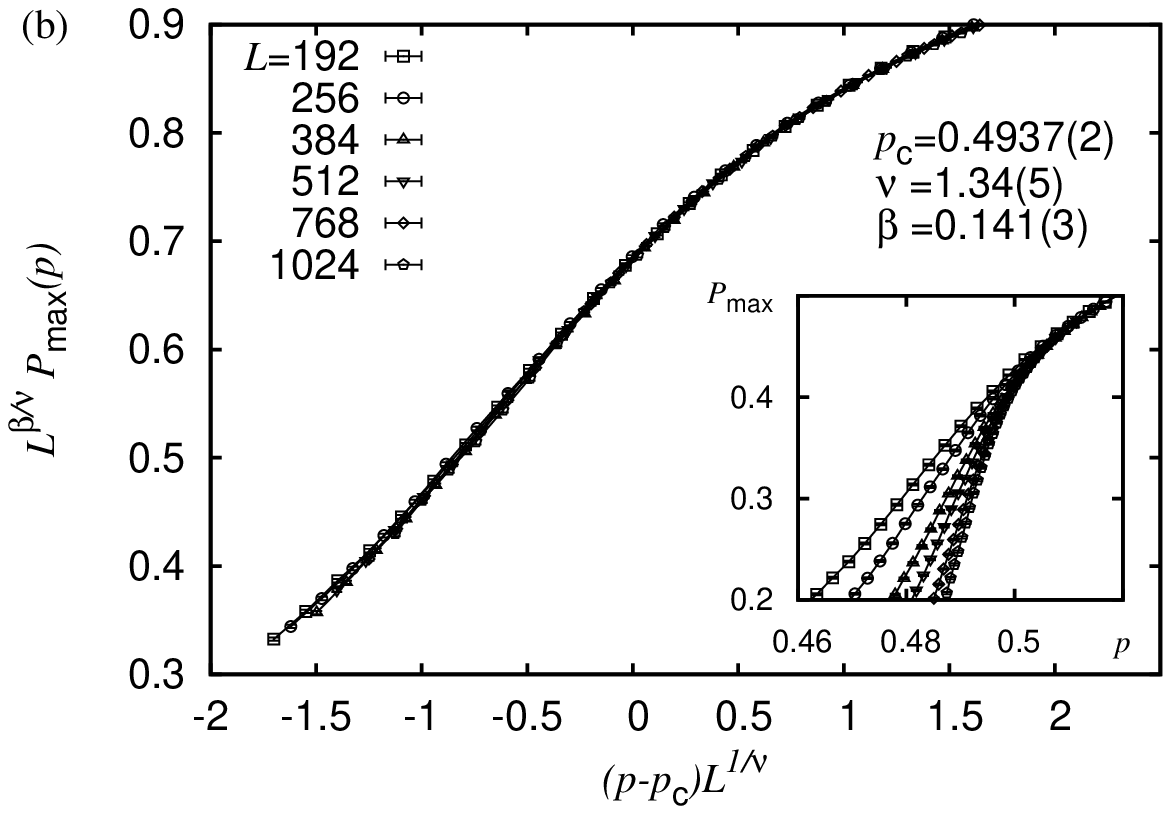}
\includegraphics[width=1.0\linewidth]{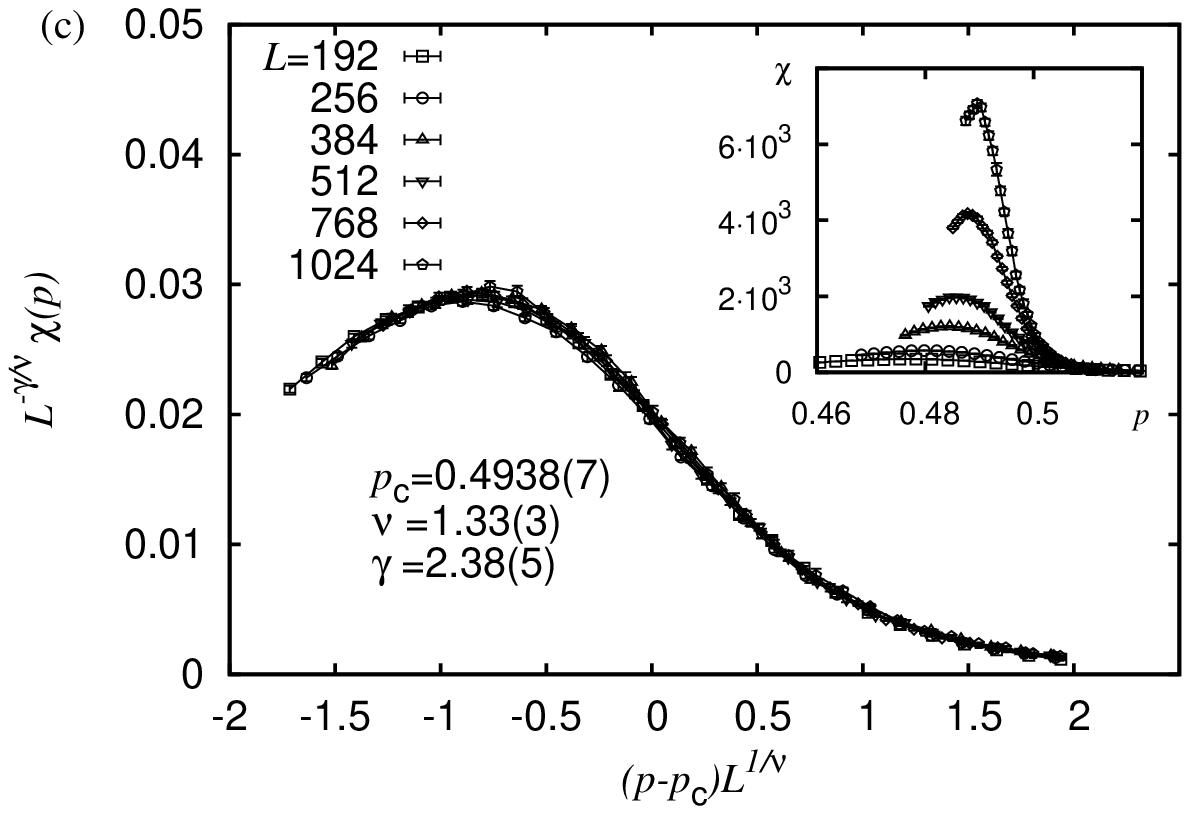}
\end{center}
\caption{Finite size scaling analyses related to the 
relative size $s_{\rm max}$ of the largest cluster 
for the CGP-R model on $2D$ square lattice of side length $L=192$ 
through $1024$, averaged over different initial seed configurations.
The main plots show the data collapse obtained according
to Eq.\ \ref{eq:scalingAssumption}, and the insets illustrate
the raw data close to the critical point.
The subfigures show different ways to analyze $s_{\rm max}$
in terms of
(a) the Binder parameter $b(p)$,
(b) the order parameter $s_{\rm max}(p)$ itself, and,
(c) the fluctuation $\chi(p)=N{\rm var}(s_{\rm max})$ 
of the order parameter.
\label{fig:orderPar_fss}}
\end{figure}

\paragraph{Order parameter:}
As a second observable we consider the order parameter, given by the relative size 
\begin{eqnarray}
P_{\rm max}(p)=\langle s_{\rm max}(p) \rangle \label{eq:orderPar}
\end{eqnarray}
of the largest cluster
in the final configuration, averaged over different initial seed configurations.
This observable scales according to Eq.\ (\ref{eq:scalingAssumption}), where $a=1/\nu$ and
$b=-\beta/\nu$. Again, a data collapse for the four largest system sizes in the range
$\epsilon \in [-2,2]$ yields the scaling parameters $p_c=0.4937(2)$, $\nu=1.34(5)$, and
$\beta=0.141(3)$ for a quality $S=0.77$ of the data collapse, see Fig.\ \ref{fig:orderPar_fss}(b).
Note that the numerical value of the order parameter exponent $\beta$ is in agreement with the 
standard percolation exponent $\beta_{\rm perc}=5/36\approx 0.139$ (Ref.\ \cite{Tsakiris2010b} reports $\beta=0.24$). 

A different way to analyze the same set of data is in terms of the Binder ratio \cite{binder1981}
\begin{eqnarray}
b(p) = \frac{1}{2} \Big[3 - \frac{\langle s_{\rm max}^4(p) \rangle}{ \langle s_{\rm max}^2(p) \rangle^2} \Big].   \label{eq:binderPar}
\end{eqnarray}
This observable scales according to Eq.\ (\ref{eq:scalingAssumption}), where, as for the spanning probability above, $a=1/\nu$ and $b=0$.
The best data collapse yields $p_c=0.4937(4)$, and $\nu=1.34(8)$ with a quality $S=0.96$ in the 
range $\epsilon \in [-2,2]$, see Fig.\ \ref{fig:orderPar_fss}(a).

A further critical exponent can be estimated from the scaling of the order parameter fluctuations $\chi(p)$, defined as
\begin{eqnarray}
\chi(p)= N [ \langle s_{\rm max}^2(p) \rangle - \langle s_{\rm max}(p) \rangle^2  ]. \label{eq:suscept}
\end{eqnarray}
The order parameter fluctuations are expected to scale according to Eq.\ (\ref{eq:scalingAssumption}), 
where $a=1/\nu$, and $b=-\gamma/\nu$.  
A best data collapse is attained for $p_c=0.4938(7)$, $\nu=1.33(3)$, and $\gamma=2.38(5)$ with a quality $S=0.84$, again 
in the range $\epsilon \in [-2,2]$, see Fig.\ \ref{fig:orderPar_fss}(c).
Here, the numerical value of the fluctuation exponent $\gamma$ is in agreement with the 
standard percolation exponent $\gamma_{\rm perc}=43/18\approx 2.389$ (Ref.\ \cite{Tsakiris2010b} reports $\gamma=1.91$). 

\paragraph{Average size of the finite clusters:}
As a third observable we consider the average size $\langle S_{\rm fin} (p)\rangle$ of all finite clusters for a particular 
final configuration, averaged over different initial seed configurations \cite{Sur1976,stauffer1994}.
This observable is defined as 
\begin{eqnarray}
S_{\rm fin}(p) = \frac{\sum_{s}^\prime s^2\, n_s(p)}{\sum_{s}^\prime s\, n_s(p)}, \label{eq:finClusters}
\end{eqnarray}
where $n_s(p)$ indicates the probability mass function of cluster sizes for a single sample 
at a given value of $p$. Note that the sums run over finite clusters only \cite{Sur1976,stauffer1994} (indicated by the prime), i.e.\ if a final 
configuration features a system spanning cluster, this cluster is excluded from the sums.
The average size of all finite clusters is expected to scale according to Eq.\ (\ref{eq:scalingAssumption}),
where $a=1/\nu$ and $b=-\gamma/\nu$.
For this observable, considering the system sizes $L=128, 192, 256, 384$, 
a best data collapse is found for $p_c=0.491(1)$, $\nu=1.35(2)$, and $\gamma=2.37(3)$
with a quality $S=1.81$ of the data collapse. 
Considering the sequence of larger system sizes $L=384, 512, 768, 1024$, 
the optimal choices for the scaling parameters result in $p_c=0.4930(5)$, $\nu=1.35(2)$, and $\gamma=2.35(4)$
with a quality $S=1.16$.
Both results agree within errorbars and were obtained in the range $\epsilon \in [-1,1]$. 
Further, the value of $\gamma$ estimated from the average size of the finite clusters support 
the numerical value estimated from the scaling behavior of the order parameter fluctuations.

\begin{figure}[t!]
\begin{center}
\includegraphics[width=1.0\linewidth]{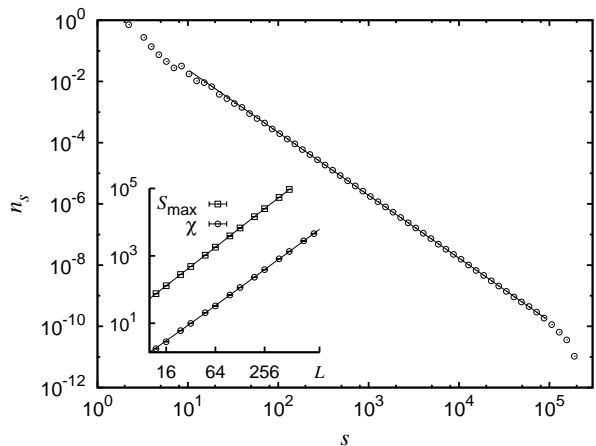}
\end{center}
\caption{Scaling analysis of different observables
at the critical point $p_c$ of the CGP-R model. The main plot shows the 
probability mass function $n_s$ of clusters of size $s$, 
displayed using logarithmic binning.
The inset illustrates the scaling of the average size $\langle S_{\rm max} \rangle$
of the largest cluster and the order parameter fluctuations $\chi$.
\label{fig:scaling_pc}}
\end{figure}

\paragraph{Further observables recorded at $p_c$:}
The various estimates for the critical point are in agreement with $p_c=0.494(1)$. At this 
critical point, Eq.\ (\ref{eq:scalingAssumption}) reduces to $y(p_c,L)\propto L^{-b}$. 
We performed additional simulations at $p_c$ to determine the scaling dimension of 
the largest cluster $S_{\rm max}$, defining the fractal dimension $d_f$ according 
to $\langle S_{\rm max} \rangle \propto L^{d_f}$. 
Considering the scaling form $\langle S_{\rm max} \rangle(L)=a L^{d_f}$ 
for a fit in the interval $L\in [20,1024]$, 
we find $a=0.669(1)$, and $d_f=1.8995(4)$ (where the reduced $\chi^2$-value reads $\chi^2_{\rm red}=0.65$), 
see inset of Fig.\ \ref{fig:scaling_pc}.
Albeit the error bars obtained from the least-squares fit is notoriously small, the numerical 
value of the fractal dimension itself is in good agreement with the ordinary percolation
estimate $d_{f, {\rm perc}}=91/48\approx 1.896$
(Ref.\ \cite{Tsakiris2010b} report $d_f=1.79(1)$).

For numerical redundancy we further estimate the exponent ratio $\gamma/\nu$ from 
the order parameter fluctuations, Eq.\ (\ref{eq:suscept}), at $p_c$ via
$\chi(L) \propto L^{\gamma/\nu}$, where a fit to the interval $L\in [20,1024]$ yields $\gamma/\nu=1.772(3)$ ($\chi^2_{\rm red}=1.0$).
Allowing for a slight deviation from the pure power law form using the effective scaling
form $\chi(L) \propto (L+\Delta L)^{\gamma/\nu}$, we obtain
$\gamma/\nu=1.785(8)$ ($\chi^2_{\rm red}=0.81$), see inset of  Fig.\ \ref{fig:scaling_pc}.
Note that the resulting exponent ratio is not only in good agreement with the numerical values for 
$\gamma$ and $\nu$ obtained above, but also with the standard percolation estimate $\gamma_{\rm perc}/\nu_{\rm perc}=129/72 \approx 1.792$
(Ref.\ \cite{Tsakiris2010b} report $\gamma/\nu=1.63(1)$).

Further, at the critical point we recorded the probability mass function $n_s$ for clusters
of size $s$. It exhibits an algebraic decay following $n_s \propto s^{-\tau}$, for which we estimate
the exponent $\tau=2.057(1)$. This estimate was obtained for the data recorded at $L=1024$ using a 
least-squares fit to the interval $s\in [200,10000]$ (the errorbar was estimate by bootstrap 
resampling of the underlying datasets), see main plot of Fig.\ \ref{fig:scaling_pc}. Note that the value is reasonably close to the standard percolation
estimate $\tau_{\rm perc}=187/91\approx 2.055$ (Ref.\ \cite{Tsakiris2010b} report $\tau=2.08$, computed using a hyperscaling relation).

Finally, at $p_c$ and on a large lattice of $L=1024$ we estimate the fraction of occupied sites
in the final configuration to be $p_{\rm fin}=0.589697(5)$ (for a smaller lattice setup with $L=128$ 
we find $p_{\rm fin}=0.58964(4)$). These values are only slightly smaller than the fraction $p_{c, {\rm perc}}=0.59274621(13)$ of 
occupied sites at the critical point of $2D$ site percolation \cite{Ziff2000,wiki:PercThreshold}.

\section{Conclusions \label{sect:conclusions}}

We have investigated the continuous transition of the ``Touch and Stop''
cluster growth percolation model at intermediate values of the initial 
fraction $p$ of seed sites by means of numerical simulations, implementing
rhombic and disk-shaped cluster growth.
Previously, the critical point and the critical exponents that 
govern the transition were estimated \cite{Tsakiris2010b} and it was concluded that
the transition is in a different universality class than standard 
percolation. 
From a point of view of data analysis, the previously presented analysis 
could be improved in various regards.
Here, we have revisited the cluster growth percolation model 
and performed a finite-size scaling analysis considering the
geometric properties of the largest clusters for the individual 
final configurations attained after the growth procedure is completed.
Using large system sizes and appropriate sample sizes, we obtained
highly precise estimates for the critical points of both cluster growth styles 
and the usual critical 
exponents that characterize the scaling behavior of the spanning probability, 
the order parameter (and its fluctuations), and the probability mass function
of cluster sizes right at the critical point.
We further used different observables to estimate individual exponents.
E.g., in order to determine the fluctuation exponent $\gamma$, we considered
the order parameter fluctuations $\chi$ (using the data collapse techniques
shown in Fig.\ \ref{fig:orderPar_fss}(c) as well as 
the finite-size scaling right at $p_c$ illustrated in the inset of Fig.\ \ref{fig:scaling_pc}) 
and the average size of the finite clusters $S_{\rm fin}$. 
This lead to highly redundant numerical estimates for the 
critical exponents and yields maximally justifiable results.

In summary, we find that the critical exponents estimated for the ``Touch and Stop''
cluster growth model are in reasonable agreement with those that describe the 
standard $2D$ percolation phenomenon, see Tab.\ \ref{tab:tab1}. 
Hence we conclude that the $2D$ cluster growth 
percolation model is in the same universality class as standard $2D$ percolation.

Further, we found that in the vicinity of the critical point, the cluster growth procedure 
(for $L>256$) is completed after $3-4$ steps. I.e., in comparison to ordinary site percolation,
the growth procedure can effectively be seen as a process that affects the cluster 
configurations close to $p_c$ only locally. 
This might render it somewhat intuitive that the cluster growth procedure leads to a 
shift of the (anyway non-universal) critical point but does not change 
the critical exponents as compared to standard percolation.

\begin{acknowledgments}
I thank M.\ Nartey for a contribution to the ``Paperclub'' of 
the Computational theoretical physics group at the University of Oldenburg, 
which motivated this study.
Further, I am much obliged to R.\ Ziff, A.~K.\ Hartmann, P.\ Argyrakis and
C.\ Norrenbrock for valuable discussions and for commenting on an early 
draft of the manuscript.
Finally, I gratefully acknowledge financial support from the DFG 
(\emph{Deutsche Forschungsgemeinschaft}) under grant HA3169/3-1.
The simulations were performed at the HPC Cluster HERO, located at 
the University of Oldenburg (Germany) and funded by the DFG through
its Major Instrumentation Programme (INST 184/108-1 FUGG) and the
Ministry of Science and Culture (MWK) of the Lower Saxony State.
\end{acknowledgments}


\appendix

\begin{figure}[t!]
\begin{center}
\includegraphics[width=1.0\linewidth]{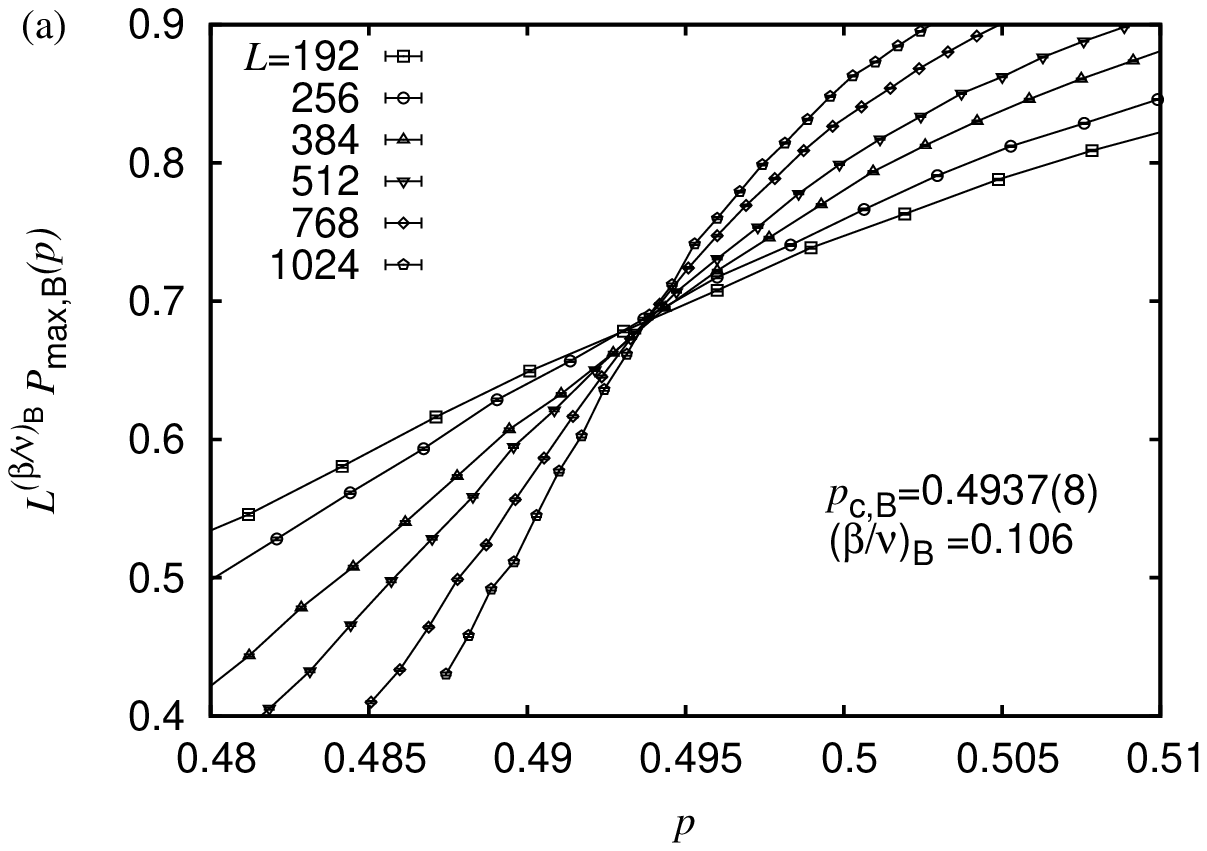}
\includegraphics[width=1.0\linewidth]{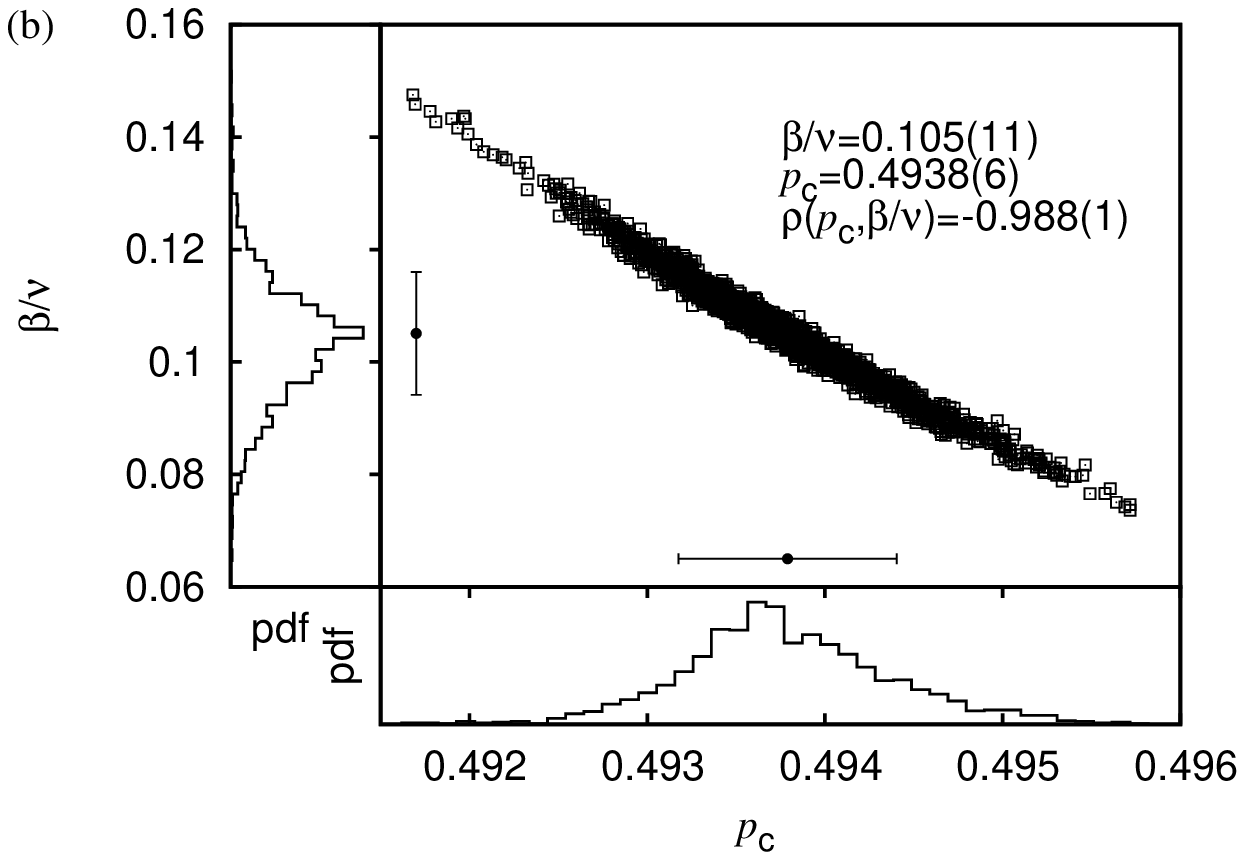}
\end{center}
\caption{Order parameter analysis for the CGP-R model, similar to that
performed by Tsakiris \emph{et al}.
(a) relative size $s_{\rm max}$ of the largest
cluster for different system sizes scaled by 
a factor of $L^{\beta/\nu}$, where $\beta/\nu$ is 
adjusted such that the data curves exhibit a 
common crossing point (see text for more details).
(b) results of a bootstrap resampling analysis for 
the analysis in (a). The central figure shows the distribution
of tuples $(\beta/\nu,p_{\rm c})$  in the plane and 
the small adjacent plots show the probability density 
function along the independent directions. The points with error bars 
in the central figure show the resulting estimates for
the corresponding resampled parameters.
\label{fig:orderPar_argyrakis}}
\end{figure}  

\section{Order parameter analysis similar to that of Tsakiris \emph{et al.} \label{ap:appendixA}}

An admissible criticism of the analysis presented earlier in 
sect.\ \ref{sect:results} is, by considering the order-parameter 
and using the data-collapse method, one has in principle three 
parameters ($p_c$, $\nu$ and $\beta$) to adjust in order to yield a 
data collapse.
From a practical point of view, one might first consider a 
dimensionless quantity as, e.g., the percolation probability or the 
binder ratio, to arrive at estimates for $p_c$ and $\nu$ (which are 
still two parameters to adjust at once). These can then be inserted into the 
scaling relation for the order-parameter, leaving only the scaling 
parameter $\beta$ left to adjust in order to obtain a best collapse 
of the order-parameter curves. We found that whatever protocol we followed, 
the numerical estimates of the scaling parameters did vary only within their 
respective errorbars. Hence, from a point of view of analysis using 
the data-collapse method, the numerical estimates of $p_c$, $\nu$,
and $\beta$ (as listed in Tab.\ \ref{tab:tab1}) appear to be consistent and reliable.

However, we also performed an analysis of the order-parameter similar to that 
of Tsakiris \emph{et al.}, cf.\ Fig.\ (4) of Ref.\ \cite{Tsakiris2010b}.
The benefit of their ``crossing-point'' method is that it features only 
one scaling parameter, namely the exponent ratio $\beta/\nu$ \cite{ArgyrakisPrivComm}. 
In the remainder of this appendix we explain the ``crossing-point'' analysis 
procedure and report on the results obtained therewith.

Starting with the data curves for the relative size $s_{\rm max}$ of 
the largest cluster for different system sizes, we scale the 
data curves by a factor of of $L^{\beta/\nu}$.
If we consider 
$N$ data curves subject to $\beta/\nu>0$, we yield $M=(N(N-1))/2$ 
presumably different crossing points $p_i$, $i=1,\ldots,M$.
The exponent ratio $\beta/\nu$ is then adjusted in order to minimize 
the width $\Delta p = {\rm max}(p_i)-{\rm min}(p_i)$ of the set of
crossing points, resulting in a tuple $(\beta/\nu,p_{\rm c})$ of ``optimal'' parameters
(we set $p_c=\sum_i p_i/M$).
In order to estimate errors for these optimal parameters, we performed a
bootstrap resampling procedure of the raw data and analyzed each resampled
set of data curves using the above procedure. We considered an overall number
of $N=6$ data curves and $m=2000$ bootstrap samples. Fig.\ \ref{fig:orderPar_argyrakis}(a) shows one 
of these bootstrap samples.
The analysis of all $m$ optimal tuples $(\beta/\nu,p_{\rm c})$ in the plane is 
summarized in Fig.\ \ref{fig:orderPar_argyrakis}(b).
Therein, the central plot shows the distribution of tuples in the plane and
the small adjacent plots show the probability density 
function along the independent directions. 
The resulting estimates for the resampled optimal parameters read 
$\beta/\nu=0.105(11)$ and $p_{\rm c}=0.4937(8)$. Note that these
are in excellent agreement with the results obtained from the 
data collapse analysis, listed in Tab.\ \ref{tab:tab1}. Further, 
Pearson's correlation coefficient between the two parameters reads 
$\rho(\beta/\nu,p_{\rm c})=-0.988(1)$. This indicates that for 
an increasing numerical value of the exponent ratio, the optimized 
crossing point shifts to smaller values of $p$.
However, note that the above analysis does not account for systematic
deviations from scaling. Similar to conventional data-collapse techniques
it analyses the data ``as observed'', implying that the scaling
assumption holds.

\begin{figure}[t!]
\centerline{
\includegraphics[width=1.0\linewidth]{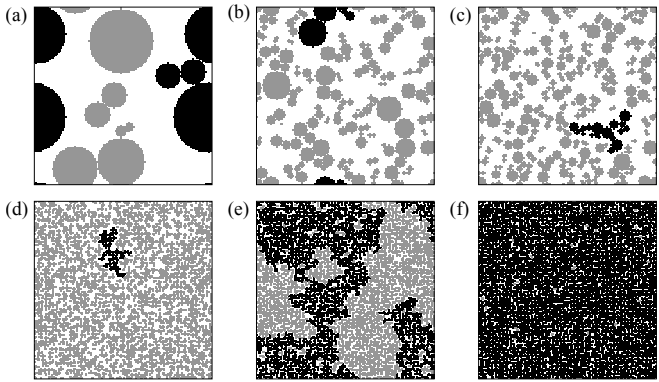}}
\caption{
Samples of final configurations obtained for the CGP-D model 
on $2D$ square lattices of side length $L\!=\!128$ using 
disk-shaped cluster growth.
The snapshots relate to different values of the
initial density of active seed sites:
$p\!=\!0.001$, $0.01$, $0.02$, $0.2$, $0.495$, $0.7$, for 
subfigures (a--f), respectively. 
In the limit of large system sizes and above the critical point 
$p_c\approx 0.497$, the largest cluster in the final configuration spans the 
lattice along at least one direction. In each subfigure, the largest
cluster is colored black and all finite clusters are colored grey.
\label{fig:2Dsamples_circles}}
\end{figure}  

\section{Results for disk-shaped cluster growth \label{ap:appendixB}}
Assuming a circular shape of the growing clusters (see Fig.\ \ref{fig:2Dsamples_circles}), 
and upon analysis of the
order parameter (again for square systems with side-length up to $L=1024$), we yield the following results for the critical properties 
of the $2D$ ``Touch and stop'' cluster growth model:

\paragraph{Binder ratio:}
Considering the Binder ratio, the best data collapse (obtained in the range $\epsilon \in [-2,2]$)
yields $p_c=0.4978(3)$, and $\nu=1.32(4)$ with a quality $S=0.984$.

\paragraph{Order parameter:}
Considering the relative size of the largest cluster found in the final configuration, the 
best data collapse (obtained in the range $\epsilon \in [-1,1]$)
yields $p_c=0.4974(1)$, $\nu=1.29(3)$, and $\beta=0.145(4)$ with a quality $S=0.29$.

\paragraph{Order parameter fluctuations:}
A best data collapse for the order parameter fluctuations (attained in the range $\epsilon \in [-1.5,1.5]$) 
results in the estimates $p_c=0.4978(5)$, $\nu=1.33(2)$, and $\gamma=2.38(3)$ with a quality $S=0.89$.

\paragraph{Cluster size distribution:}
At the critical point,
the algebraic decay of the probability mass function $n_s$ for clusters
of size $s$ is governed by the exponent $\tau=2.050(6)$. 
This estimate was obtained for the data recorded at $L=1024$, using a 
least-squares fit to the interval $s\in [500,7000]$ (as before, the errorbar was estimated 
via bootstrap resampling).

The observation that the critical points for rhombic and disk-shaped cluster
growth differ only slightly can be understood from the time evolution of 
the individual clusters.
At short times (i.e.\ $t\leq2$) both ``growth styles'' 
feature identical clusters. Only at times $t>2$, the disk-shaped
growth style leads to more convex clusters that assume a 
circular shape in the limit $t\to\infty$.
Close to the critical points of both growth styles, the cluster
growth procedure stops after maximally $4$ time steps (i.e.\ at $t=4$).
Thereby, the majority of initial seeds grow for less than
4 time steps and become inactive in a state where one cannot tell apart
clusters that were grown using a rhombic or disk-shaped growth rule.
Note that this is due to the discreteness of the underlying lattice.


\bibliography{cgp_fss.bib}

\begin{thebibliography}{17}
\expandafter\ifx\csname natexlab\endcsname\relax\def\natexlab#1{#1}\fi
\expandafter\ifx\csname bibnamefont\endcsname\relax
  \def\bibnamefont#1{#1}\fi
\expandafter\ifx\csname bibfnamefont\endcsname\relax
  \def\bibfnamefont#1{#1}\fi
\expandafter\ifx\csname citenamefont\endcsname\relax
  \def\citenamefont#1{#1}\fi
\expandafter\ifx\csname url\endcsname\relax
  \def\url#1{\texttt{#1}}\fi
\expandafter\ifx\csname urlprefix\endcsname\relax\def\urlprefix{URL }\fi
\providecommand{\bibinfo}[2]{#2}
\providecommand{\eprint}[2][]{\url{#2}}

\bibitem[{\citenamefont{Stauffer}(1979)}]{stauffer1979}
\bibinfo{author}{\bibfnamefont{D.}~\bibnamefont{Stauffer}},
  \bibinfo{journal}{Phys. Rep.} \textbf{\bibinfo{volume}{54}},
  \bibinfo{pages}{1} (\bibinfo{year}{1979}).

\bibitem[{\citenamefont{Stauffer and Aharony}(1994)}]{stauffer1994}
\bibinfo{author}{\bibfnamefont{D.}~\bibnamefont{Stauffer}} \bibnamefont{and}
  \bibinfo{author}{\bibfnamefont{A.}~\bibnamefont{Aharony}},
  \emph{\bibinfo{title}{{Introduction to Percolation Theory}}}
  (\bibinfo{publisher}{Taylor and Francis, London}, \bibinfo{year}{1994}).

\bibitem[{\citenamefont{Pfeiffer and Rieger}(2002)}]{pfeiffer2002}
\bibinfo{author}{\bibfnamefont{F.~O.} \bibnamefont{Pfeiffer}} \bibnamefont{and}
  \bibinfo{author}{\bibfnamefont{H.}~\bibnamefont{Rieger}},
  \bibinfo{journal}{J. Phys.: Condens. Matter} \textbf{\bibinfo{volume}{14}},
  \bibinfo{pages}{2361} (\bibinfo{year}{2002}).

\bibitem[{\citenamefont{Pfeiffer and Rieger}(2003)}]{pfeiffer2003}
\bibinfo{author}{\bibfnamefont{F.~O.} \bibnamefont{Pfeiffer}} \bibnamefont{and}
  \bibinfo{author}{\bibfnamefont{H.}~\bibnamefont{Rieger}},
  \bibinfo{journal}{Phys. Rev. {\bf E}} \textbf{\bibinfo{volume}{67}},
  \bibinfo{pages}{056113} (\bibinfo{year}{2003}).

\bibitem[{\citenamefont{Cieplak et~al.}(1994)\citenamefont{Cieplak, Maritan,
  and Banavar}}]{cieplak1994}
\bibinfo{author}{\bibfnamefont{M.}~\bibnamefont{Cieplak}},
  \bibinfo{author}{\bibfnamefont{A.}~\bibnamefont{Maritan}}, \bibnamefont{and}
  \bibinfo{author}{\bibfnamefont{J.~R.} \bibnamefont{Banavar}},
  \bibinfo{journal}{Phys. Rev. Lett.} \textbf{\bibinfo{volume}{72}},
  \bibinfo{pages}{2320} (\bibinfo{year}{1994}).

\bibitem[{\citenamefont{Melchert and Hartmann}(2007)}]{melchert2007}
\bibinfo{author}{\bibfnamefont{O.}~\bibnamefont{Melchert}} \bibnamefont{and}
  \bibinfo{author}{\bibfnamefont{A.~K.} \bibnamefont{Hartmann}},
  \bibinfo{journal}{Phys. Rev. B} \textbf{\bibinfo{volume}{76}},
  \bibinfo{pages}{174411} (\bibinfo{year}{2007}).

\bibitem[{\citenamefont{Schwarz et~al.}(2009)\citenamefont{Schwarz,
  Karrenbauer, Schehr, and Rieger}}]{schwarz2009}
\bibinfo{author}{\bibfnamefont{K.}~\bibnamefont{Schwarz}},
  \bibinfo{author}{\bibfnamefont{A.}~\bibnamefont{Karrenbauer}},
  \bibinfo{author}{\bibfnamefont{G.}~\bibnamefont{Schehr}}, \bibnamefont{and}
  \bibinfo{author}{\bibfnamefont{H.}~\bibnamefont{Rieger}},
  \bibinfo{journal}{J. Stat. Mech.} \textbf{\bibinfo{volume}{2009}},
  \bibinfo{pages}{P08022} (\bibinfo{year}{2009}).

\bibitem[{\citenamefont{Mertens and Moore}(2012)}]{MertensMoore2012}
\bibinfo{author}{\bibfnamefont{S.}~\bibnamefont{Mertens}} \bibnamefont{and}
  \bibinfo{author}{\bibfnamefont{C.}~\bibnamefont{Moore}}
  (\bibinfo{year}{2012}), \bibinfo{note}{preprint: arXiv:1209.4936}.

\bibitem[{\citenamefont{Tsakiris
  et~al.}(2010{\natexlab{a}})\citenamefont{Tsakiris, Maragakis, Kosmidis, and
  Argyrakis}}]{Tsakiris2010a}
\bibinfo{author}{\bibfnamefont{N.}~\bibnamefont{Tsakiris}},
  \bibinfo{author}{\bibfnamefont{M.}~\bibnamefont{Maragakis}},
  \bibinfo{author}{\bibfnamefont{K.}~\bibnamefont{Kosmidis}}, \bibnamefont{and}
  \bibinfo{author}{\bibfnamefont{P.}~\bibnamefont{Argyrakis}}
  (\bibinfo{year}{2010}{\natexlab{a}}), \bibinfo{note}{{preprint:
  arXiv:1004.1526v2; A summary of this article is available at papercore.org,
  see {http://www.papercore.org/Tsakiris2010a}}}.

\bibitem[{\citenamefont{Tsakiris
  et~al.}(2010{\natexlab{b}})\citenamefont{Tsakiris, Maragakis, Kosmidis, and
  Argyrakis}}]{Tsakiris2010b}
\bibinfo{author}{\bibfnamefont{N.}~\bibnamefont{Tsakiris}},
  \bibinfo{author}{\bibfnamefont{M.}~\bibnamefont{Maragakis}},
  \bibinfo{author}{\bibfnamefont{K.}~\bibnamefont{Kosmidis}}, \bibnamefont{and}
  \bibinfo{author}{\bibfnamefont{P.}~\bibnamefont{Argyrakis}}
  (\bibinfo{year}{2010}{\natexlab{b}}), \bibinfo{note}{{preprint:
  arXiv:1004.5028; A summary of this article is available at papercore.org, see
  {http://www.papercore.org/Tsakiris2010}}}.

\bibitem[{\citenamefont{Houdayer and Hartmann}(2004)}]{houdayer2004}
\bibinfo{author}{\bibfnamefont{J.}~\bibnamefont{Houdayer}} \bibnamefont{and}
  \bibinfo{author}{\bibfnamefont{A.~K.} \bibnamefont{Hartmann}},
  \bibinfo{journal}{Phys. Rev. {\bf B}} \textbf{\bibinfo{volume}{70}},
  \bibinfo{pages}{014418} (\bibinfo{year}{2004}).

\bibitem[{\citenamefont{Melchert}(2009)}]{autoScale2009}
\bibinfo{author}{\bibfnamefont{O.}~\bibnamefont{Melchert}},
  \bibinfo{journal}{Preprint: arXiv:0910.5403v1}  (\bibinfo{year}{2009}).

\bibitem[{\citenamefont{Binder}(1981)}]{binder1981}
\bibinfo{author}{\bibfnamefont{K.}~\bibnamefont{Binder}}, \bibinfo{journal}{Z.
  Phys. B} \textbf{\bibinfo{volume}{43}}, \bibinfo{pages}{119}
  (\bibinfo{year}{1981}).

\bibitem[{\citenamefont{Sur et~al.}(1976)\citenamefont{Sur, Lebowitz, Marro,
  Kalos, and Kirkpatrick}}]{Sur1976}
\bibinfo{author}{\bibfnamefont{A.}~\bibnamefont{Sur}},
  \bibinfo{author}{\bibfnamefont{J.~L.} \bibnamefont{Lebowitz}},
  \bibinfo{author}{\bibfnamefont{J.}~\bibnamefont{Marro}},
  \bibinfo{author}{\bibfnamefont{M.~H.} \bibnamefont{Kalos}}, \bibnamefont{and}
  \bibinfo{author}{\bibfnamefont{S.}~\bibnamefont{Kirkpatrick}},
  \bibinfo{journal}{J. Stat. Phys.} \textbf{\bibinfo{volume}{15}},
  \bibinfo{pages}{345} (\bibinfo{year}{1976}).

\bibitem[{\citenamefont{Newman and Ziff}(2000)}]{Ziff2000}
\bibinfo{author}{\bibfnamefont{M.}~\bibnamefont{Newman}} \bibnamefont{and}
  \bibinfo{author}{\bibfnamefont{R.}~\bibnamefont{Ziff}},
  \bibinfo{journal}{Phys. Rev. Lett.} \textbf{\bibinfo{volume}{85}},
  \bibinfo{pages}{4104} (\bibinfo{year}{2000}), \bibinfo{note}{{A summary of
  this article is available at papercore.org, see
  {http://www.papercore.org/Newman2000}}}.

\bibitem[{\citenamefont{Wikipedia}(2012)}]{wiki:PercThreshold}
\bibinfo{author}{\bibnamefont{Wikipedia}}, \emph{\bibinfo{title}{Percolation
  threshold --- {W}ikipedia{,} the free encyclopedia}} (\bibinfo{year}{2012}),
  \bibinfo{note}{{online; accessed 27-August-2012}},
  \urlprefix\url{http://en.wikipedia.org/wiki/Percolation_threshold}.

\bibitem[{\citenamefont{Argyrakis}(2012)}]{ArgyrakisPrivComm}
\bibinfo{author}{\bibfnamefont{P.}~\bibnamefont{Argyrakis}}
  (\bibinfo{year}{2012}), \bibinfo{note}{private communication}.

\end{thebibliography}

\end{document}